\documentclass[aps,pra,reprint,superscriptaddress]{revtex4-2}

\usepackage[english]{babel}
\usepackage[utf8]{inputenc}
\usepackage[top=2cm, bottom=2cm, left=2cm, right=2cm]{geometry}
\usepackage{color,soul}
\usepackage{graphicx}
\usepackage{siunitx}

\usepackage[svgnames]{xcolor}
\usepackage[intlimits]{mathtools}
\usepackage{tabulary}

\usepackage[euler]{textgreek}

\usepackage{newtxmath}

\begin{document}

\title{Post-collision interaction in the liquid phase: the role of electron scattering}

\author{R\'emi Dupuy}
\affiliation{Sorbonne Université, CNRS, Laboratoire de Chimie Physique - Matière et Rayonnement, LCPMR, F-75005 Paris Cedex 05, France}
\author{Francis Penent}
\affiliation{Sorbonne Université, CNRS, Laboratoire de Chimie Physique - Matière et Rayonnement, LCPMR, F-75005 Paris Cedex 05, France}
\author{Bastien Lutet-Toti}
\affiliation{Sorbonne Université, CNRS, Laboratoire de Chimie Physique - Matière et Rayonnement, LCPMR, F-75005 Paris Cedex 05, France}
\author{Vincent Duval}
\affiliation{Sorbonne Université, CNRS, Laboratoire de Chimie Physique - Matière et Rayonnement, LCPMR, F-75005 Paris Cedex 05, France}

\author{J\'erome Palaudoux}
\affiliation{Sorbonne Université, CNRS, Laboratoire de Chimie Physique - Matière et Rayonnement, LCPMR, F-75005 Paris Cedex 05, France}
\author{Denis C\'eolin}
\affiliation{Synchrotron SOLEIL, L'Orme des Merisiers, Saint-Aubin - BP 48 91192, Gif-sur-Yvette Cedex, France}

\begin{abstract}
We report a detailed study of post-collision interaction (PCI) in a liquid medium. We investigate PCI for the Auger KLL electrons of solvated Cl$^-$, K$^+$ and Ca$^{2+}$. All three isoelectronic ions exhibit a very similar PCI behavior, which is little affected by changing the solvent from water to methanol or ethanol. The two main factors modifying the PCI interaction in condensed media is the screening of Coulombic interactions, and scattering of the electrons. The experimental results are compared with the predictions of a previously reported semi-classical PCI theory modified to account for screening and scattering. We show that a better agreement with experiment can be obtained by instead modeling scattering using electron transport Monte-Carlo simulations. We suggest that, in turn, PCI experimental data could be used as another experimental constraint to refine the currently insufficiently well-known scattering parameters of low-energy electrons in water, which are crucial in many fields. 
\end{abstract}

\maketitle

\section{Introduction}

Photoionization of inner-shells can lead to the consecutive emission of several electrons, e.g. one photoelectron and one Auger electron in the most simple case. These two electrons can exchange energy after their emission, in particular if the first one is much slower than the second and can thus be rapidly overtaken by it. In a simplified picture, after Auger emission, the change of ion charge changes the Coulomb field in which the photoelectron evolves. This change is initially screened by the Auger electron, and is felt by the photoelectron only when it is overtaken by the Auger electron. An energy corresponding to this change of field of 1/R (in atomic units, as used throughout this manuscript), where R is the crossing radius, is exchanged between the two electrons: the photoelectron is slowed down and the Auger electron accelerated. This phenomenon is well-known as post-collision interaction (PCI) in atomic physics, and leads to a shift and a characteristic asymmetric distortion of electron spectra. It has been continuously studied in the last fifty years for free atoms and molecules (e.g. \cite{russek1986,vanderstraten1988,schmidt1987,kuchiev1989,guillemin2015,lablanquie2012,gerchikov2023} and references therein). 

Although studies are more scattered, the PCI effect has also been investigated in a variety of condensed phase contexts: free and supported clusters \cite{lindblad2005,peters2010}, adsorbed atoms on surfaces \cite{kassuhlke1998}, and bulk metals \cite{miller1984,jach1981,coward2000}, insulators \cite{bahl1979,chiang1980} and polymers \cite{velasquez2023}. The occurrence of PCI in liquids, in the same system investigated here (K$^+$ and Cl$^-$ KLL decay), was also mentioned by us in a previous publication \cite{miteva2018}. It was recognized early on that an important difference between gas and condensed phase is the presence of screening effects in the latter medium, damping the Coulombic interactions and thus reducing the magnitude of the PCI effect. In metals, charge screening by conduction band electrons is very fast and thus PCI is efficiently quenched \cite{miller1984}. The onset of this metallic screening was observed in supported gold clusters \cite{peters2010}. This efficient screening is one reason why PCI is largely neglected in practical condensed phase XPS works for determination of electron binding energies. 

In dielectric media, like water, Coulomb interaction damping is described by the dielectric permittivity of the medium $\epsilon$. Lindblad et al. \cite{lindblad2005} were the first to propose a modification of the free-atom theory of PCI  (presented e.g. by Van der Straten et al. \cite{vanderstraten1988}) where the energy exchange between the two electrons is simply divided by the dielectric permittivity to account for screening. The modified theory accounted well for their experimental results on free Argon clusters. 

The other important aspect of condensed phase PCI is the influence of electron scattering. Both elastic and inelastic scattering will increase the magnitude of PCI by reducing how fast the photoelectron moves away from the ion: inelastic scattering directly reduces its speed, while elastic scattering leads to non-straight trajectories. Scattering has been little considered in previous condensed phase PCI studies; in Lindblad et al. \cite{lindblad2005} it is mentioned but not taken into account. Nonetheless, scattering turns out to be an important effect in bulk dielectric media: an apparent increase of magnitude of PCI compared to the gas-phase was recently reported by Velasquez et al. \cite{velasquez2023}, suggesting scattering dominates over screening. They proposed another modification of PCI theory to also account for scattering, which however has shortcomings that will be discussed later. Here we will propose another approach.  

Indeed, electron scattering is stochastic in nature, and its physical modeling is complex, particularly at low kinetic energies. Incorporating it in an analytical model, when it is possible at all, often requires gross approximations. A fruitful approach has been to use Monte-Carlo simulations instead \cite{werner2001}. In such models, a large number of electron trajectories are generated by randomly drawing inelastic and elastic scattering events (and sometimes other types of events depending on the purpose of the simulation), allowing to build statistically the sought after results (e.g. the electron kinetic energy spectrum). Some examples of the use of electron transport models are quantification and spectrum modeling in XPS \cite{smekal2005}, retrieval of native electron properties (spectral shape and angular distribution) before scattering \cite{signorell2016,luckhaus2017} and large-scale simulation of track structures and radiation damage in biological media \cite{incerti2018}. Here we use a simple electron transport model to simulate the PCI effect, yielding promising results which are limited by uncertainties on the physical modeling of PCI in condensed media, and on the electron transport parameters, i.e. the differential and integral inelastic and elastic scattering cross-sections in water. We suggest that PCI data could actually constitute an additional source of constraints for these parameters.  

\section{Methods}

\subsection{Experiments}

The experiments were performed at the GALAXIES beamline of the SOLEIL synchrotron facility, using the liquid microjet module of the beamline coupled to the HAXPES hemispherical electron analyzer \cite{ceolin2013}. For the present experiment, the beamline delivered a horizontally polarized beam at photon energies between 2.8 and 5.5 keV, with a typical bandwidth of 300 to 750 meV. Electrons are detected in the same direction as the light polarization, with analyzer slit and pass energy settings yielding an analyzer resolution of 0.5 eV.    

Series of KL$_{2,3}$L$_{2,3}$ Auger spectra were measured at different photon energies to look for the PCI effect. We do not consider other KLL Auger lines as they are weaker and broader. We did not look for the PCI effect on the photoelectron lines because it is difficult to make accurate measurements of low-energy electrons in water, due to the strong electron-scattering background \cite{malerz2021a}. We thus make the assumption that the PCI effect measured on the Auger electron reflects one to one the photoelectron behavior.

The following solutions were used: KCl 0.5M in water for the K$^+$ and Cl$^-$ in water dataset, CaCl$_2$ 0.5M in water for the Ca$^{2+}$ in water dataset, CaCl$_2$ 0.25M in methanol for the Cl$^-$ in methanol dataset, and CaCl$_2$ 0.25M in ethanol for the Cl$^-$ in ethanol dataset. We thus ensured that the concentration of a given ion was always similar (0.5M) for all measurements. All solutions were prepared by dissolving commercial salts from Sigma-Aldrich in either milliQ water or analytic grade ($>$ 99\% purity) methanol and ethanol. 

The spectra are presented as measured. To determine the PCI shift, i.e. the shift of the peak maximum compared with its value well-above threshold, the region of the main peak is fitted with an exponentially-modified Gaussian. The purpose of this fitting is not to deconvolve the spectrum into physically meaningful components, but only to pinpoint accurately the overall maximum. The presence of a background can affect peak positions, and conversely an inappropriate background subtraction with e.g. nonphysically motivated backgrounds could lead to incorrect maxima determinations. In the present case, we avoid subtracting any background on the spectra, on the basis that on the high kinetic energy side of the main peak the background is flat. The inelastic scattering tail of the main peak electrons onsets at 7 eV lower kinetic energy in water, and therefore will not affect the main peak itself.

To determine the PCI shift as a function of the photoelectron kinetic energy eKE, we measured the binding energies (BE) of the respective 1s shells. We then take eKE = h$\nu$ - BE$_{1s}$. The photon energy was calibrated by measurements of the O 1s level of water in each case. We assumed the O 1s binding energy is 538.1 eV \cite{thurmer2021}, thus from the kinetic energy of the O 1s photoelectrons we obtain the photon energy. The precision of such a method is not better than 0.2 eV, considering the uncertainty on possible variations of the O 1s binding energy across different solutions and possible variations of the liquid potential depending on the experimental conditions. It is however enough for our purpose.  

\subsection{Theoretical modelling of PCI}

\subsubsection{Free-atom theory}

As a starting point, we remind a few results of the free atom PCI theory as it has been described several decades ago already \cite{vanderstraten1988,russek1986,kuchiev1989} for the case considered here of a simple Auger transition after photoionization. The amplitude of probability for detecting a photoelectron at an energy -E relative to its nominal position (or equivalently an Auger electron at an energy E) is given by an overlap integral between the wavefunctions of the photoelectron before and after the Auger transition. In the eikonal approximation, valid a few eV above threshold, and which requires that the photoelectron kinetic energy eKE be much higher than E, the lifetime width \textGamma~and the Coulomb potential energy, this amplitude can be expressed as:   

\begin{equation}
    a(E) \propto \!\int_{0}^{+\infty}\negthickspace
    \mathrm{exp}\left(-\!\int_{0}^{+\infty}\!\left( \frac{\Gamma}{2} + i(E - S(t')) \right) dt' \right) dt
    \label{int_PCI}
\end{equation}

In this equation, t can be interpreted as the classical time when the Auger electron has been emitted and S(t) corresponds to the energy exchanged by the photoelectron and the Auger electron, which is given by: 

\begin{equation}
    S(t) = \frac{C}{v_{\mathrm{ph}}t}
\end{equation}

v$_{\mathrm{ph}}$ is the photoelectron speed, and the value of the factor C, which also depends on the photoelectron and Auger electron speeds, depends on the experimental geometry and observables considered. Since we do not look at the angular dependence of PCI, we must use the angle-integrated formula for C  \cite{vanderstraten1988}. In the experiment the light is polarized linearly in the direction of the detector and we assume the photoelectron angular distribution has an initial anisotropy factor of $\beta$ = 2, as should be the case for photoemission from an atomic 1s orbital. Then, for eKE $<$ E$_A$, the only case we will consider here, we have \cite{vanderstraten1988}

\begin{equation}
    C = 1 - \left(\frac{\mathrm{eKE}}{E_A}\right)^{1/2} - \frac{2}{5}\left(\frac{\mathrm{eKE}}{E_A}\right)^{3/2}
\end{equation}

Given this expression for S(t), equation \ref{int_PCI} yields: 

\begin{equation}
     |a(E)|^2 \propto \frac{1}{(\Gamma/2)^2 + E^2}
     \exp \left( \frac{2C}{v_{\mathrm{ph}}}\arctan\left(\frac{2E}{\Gamma}\right)\right)
     \label{straten_spec}
\end{equation}

The left part of this expression is a Lorentzian, which is asymmetrically modified by the right-side exponential function. This expression gives the full spectral shape, while in the literature it is common to focus on a single parameter which is the PCI shift, i.e. the displacement of the maximum of the peak relative to the PCI-free value. However, when comparing theory to experiment it is necessary to also take into account all sources of broadening which are not related to the core-hole lifetime. Broadening of the peak will have a large influence on the value of the observed PCI shift. While empirical formulas have been suggested before \cite{vanderstraten1988}, the simplest way to do this is to convolve the spectral shape of equation \ref{straten_spec} with the appropriate broadening peak shape (e.g. a Gaussian with a FWHM corresponding to the experimental resolution when that is the only other source of peak broadening). 

\subsubsection{Modifications of the theory in condensed phase}

As stated in the introduction, we consider here that PCI in the condensed phase can be described by modifying the free-atom theory to account for two effects of the condensed phase: screening of the Coulomb interaction, and scattering of the electrons. We discuss the limitations of this approach later. 

Screening can be accounted for through the dielectric permittivity of the medium $\epsilon$. This was first proposed by Lindblad et al. \cite{lindblad2005} and is also the approach taken by Velasquez et al. \cite{velasquez2023}. $\epsilon$ is a frequency-dependent property, as different processes occurring on different timescales contribute to it \cite{Atkins02}. PCI occurs on ultrafast timescales, of the order of the Auger decay time, i.e. femtoseconds. Therefore only the electronic part of the dielectric response should be considered - in other words, the medium does not have time to react and re-arrange molecular dipoles to fully screen the interactions. The dielectric permittivity at femtosecond timescales, i.e. optical frequencies, is then the square of the refraction index of the medium. Lindblad et al. \cite{lindblad2005} did consider the optical $\epsilon$, while Velasquez et al. \cite{velasquez2023}, on the other hand, lacking a literature value for the dielectric constant of their investigated polymer, chose to fit the value of $\epsilon$ to the experimental data. This yielded $\epsilon$ = 6.5, which would correspond to a refraction index n = 2.55, an unrealistically high value for a polymer for which typical values range around 1.5. In the present case of aqueous solutions the optical dielectric screening constant for water is well known and can be used for a more accurate modeling. The exact value does vary slightly over the optical range and with electrolyte concentration \cite{leyendekkers1977}, and an average value of $\epsilon$ = 1.8 was used throughout this work.

To account for electron scattering, Velasquez et al. remarked that in the semiclassical theory, photoelectron propagation is described by a damped wavefunction characterized by a damping lifetime $\Gamma$, and suggested that scattering was akin to an additional damping contribution $\Gamma_d$. Thus an effective $\Gamma_{\mathrm{eff}}$ = $\Gamma$ + $\Gamma_d$ is substituted in the formula. They give $\Gamma_d$ = $\hbar$v$_{\mathrm{ph}}$/$\lambda$, where $\lambda$ is the effective attenuation length (EAL) of the photoelectron, defined as the radial distance a photoelectron can travel away from its point of emission before having a probability 1/e of being inelastically scattered. Compared with the inelastic mean free path (IMFP), the EAL takes into account trajectory deviations due to elastic scattering in an approximate way. 

The modified line shape is then: 

\begin{equation}
     |a(E)|^2 \propto \frac{1}{(\Gamma_{\mathrm{eff}}/2)^2 + E^2}
     \exp\left(\frac{2C}{v_{\mathrm{ph}}\epsilon}\arctan\left(\frac{2E}{\Gamma_{\mathrm{eff}}}\right) \right)
     \label{modif_spec}
\end{equation}

And the modified PCI shift, as given in ref. \cite{velasquez2023}:

\begin{equation}
    \Delta E = \frac{\Gamma_{\mathrm{eff}}C}{2v_{\mathrm{ph}} \epsilon}
    = \frac{C}{2\epsilon}\left( \frac{\Gamma}{v_{\mathrm{ph}}} + \frac{1}{\lambda} \right)
    \label{modif_max}
\end{equation}

This approach behaves qualitatively as expected: a higher rate of inelastic or elastic scattering reduces the EAL, which increases the magnitude of PCI. On the other hand, justifying rigorously why scattering can be modeled as a damping term is difficult. The value of this damping term is large, yielding values for $\Gamma_{\mathrm{eff}}$ in the range of 2 to 4 eV depending on eKE. It is not clear whether these values correspond to experimental observables (such as lifetime broadenings), and the theory ends up predicting too broad lineshapes. Additionally, how much the magnitude of PCI is increased by inelastic scattering must be affected by how much the electron is slowed down upon an inelastic collision, i.e. the differential scattering cross section with respect to energy loss. The EAL, and therefore the $\Gamma_{\mathrm{eff}}$, does not contain any information on this aspect. Clearly, using a simply modified analytical formulation for PCI to account for scattering is difficult because of the complex nature of the process, which does not lend itself very well to analytical modeling. We attempt another approach here using Monte-Carlo simulations to model the effects of scattering. 

\subsection{Monte-Carlo electron trajectory simulations}

The principle of the simulation is the following: we simulate an electron trajectory for a photoelectron moving away from an ion in water, including scattering events. At some determined time, an Auger electron is emitted from the same ion, and we then also follow its trajectory. When the Auger electron overtakes the photoelectron, the simulation is stopped, and from the radius at which the two electrons crossed the energy exchange due to PCI is calculated. From a set of about a million such trajectories, a PCI curve is then generated. 

The electron transport part of the simulations follows the same general principles as many similar codes that have been developed elsewhere \cite{werner2001,signorell2020,schild2020}. We only consider scattering of the photoelectron: in the experiments, the detected Auger electrons are those that have not been inelastically scattered, and since Auger emission is isotropic for K-shell ionization and weakly anisotropic otherwise in most cases \cite{cleff1974,berezhko1977}, we neglect elastic scattering for the Auger electron as well. Scattering is modeled by a simplified two-channel picture: we consider electronic inelastic scattering on one side, which causes energy losses of 7 eV or more (7 eV being the approximate threshold for electronic excitation of condensed water), and elastic scattering on the other side, including quasi-elastic scattering channels such as vibrational scattering without distinction and without taking into account the related small ($<$ 0.5 eV) energy losses. From the elastic and inelastic scattering cross-sections, time steps between consecutive scattering events are drawn randomly. At each scattering event, if the event is elastic, a deflection angle is drawn from the differential elastic scattering cross-section, and the trajectory is adjusted. If it is inelastic, an energy loss is drawn from the differential inelastic scattering cross-section, and the photoelectron kinetic energy is changed in consequence, along with all the energy-dependent electron transport parameters. Between scattering events, propagation of the electrons is calculated by solving the equations of motion, based on the two-body (before Auger emission) or three-body (after Auger emission) Coulomb interactions. The main purpose of this is to account for the deceleration of the electrons as they move away from the ion. The effect of the surface is also taken into account: the point of emission is drawn at a certain distance from the surface according to the EAL of the Auger electron, and if the photoelectron leaves the liquid, scattering is turned off. 

Two different sets of electron transport parameters were used and compared for the simulation. The first one, labeled as the "ice cross-sections (CS)" set in the rest of the manuscript, makes use of the experimental elastic and inelastic scattering cross-sections measured on amorphous ice films \cite{michaud2003}, which were recently shown to reproduce sufficiently well a number of liquid-phase electron spectroscopy experiments \cite{signorell2020}. The second one, labeled as "model cross-sections", uses analytical approximations for the integral and differential cross-sections which are based on model calculations and on the better known gas phase cross-sections. It is more akin to the kind of approaches developed in models such as GEANT4-DNA \cite{incerti2018}, although with much less sophistication. Details on these parameter sets are described in the Appendix. One important point is that the integral cross-sections for both elastic and inelastic scattering are higher for the model cross-section parameter set than for the ice cross-section one; this discrepancy is discussed e.g. in ref \cite{nikjoo2016}.

To compute the PCI shift and lineshape, we used two different approaches. The first is a purely classical approach. We take the histogram of exchange energies, and convolve it with the experimental PCI-free line shape (a sum of Voigt profiles with the lorentzian core-hole width and the experimental gaussian width) which is considered to represent the initial distribution of Auger energies to be subsequently modified by PCI. As it was not clear whether this classical approach was correct, we also attempted a different approach, which retains part of the semiclassical free-atom theory. We take the table of energy exchange versus time (i.e., S(t)) obtained from the simulation, and use it to compute numerically the integral given in eq. \ref{int_PCI}. The resulting shape is convolved by a sum of two Gaussians (without the Lorentzian lifetime component which is already accounted for) for comparison with the experimental results. The validity and merits of either approaches are discussed later in the article. 

\section{Results}

\subsection{Experimental results}

\begin{figure}
    \includegraphics[trim={0cm 0cm 0cm 0cm},clip,width=\linewidth]{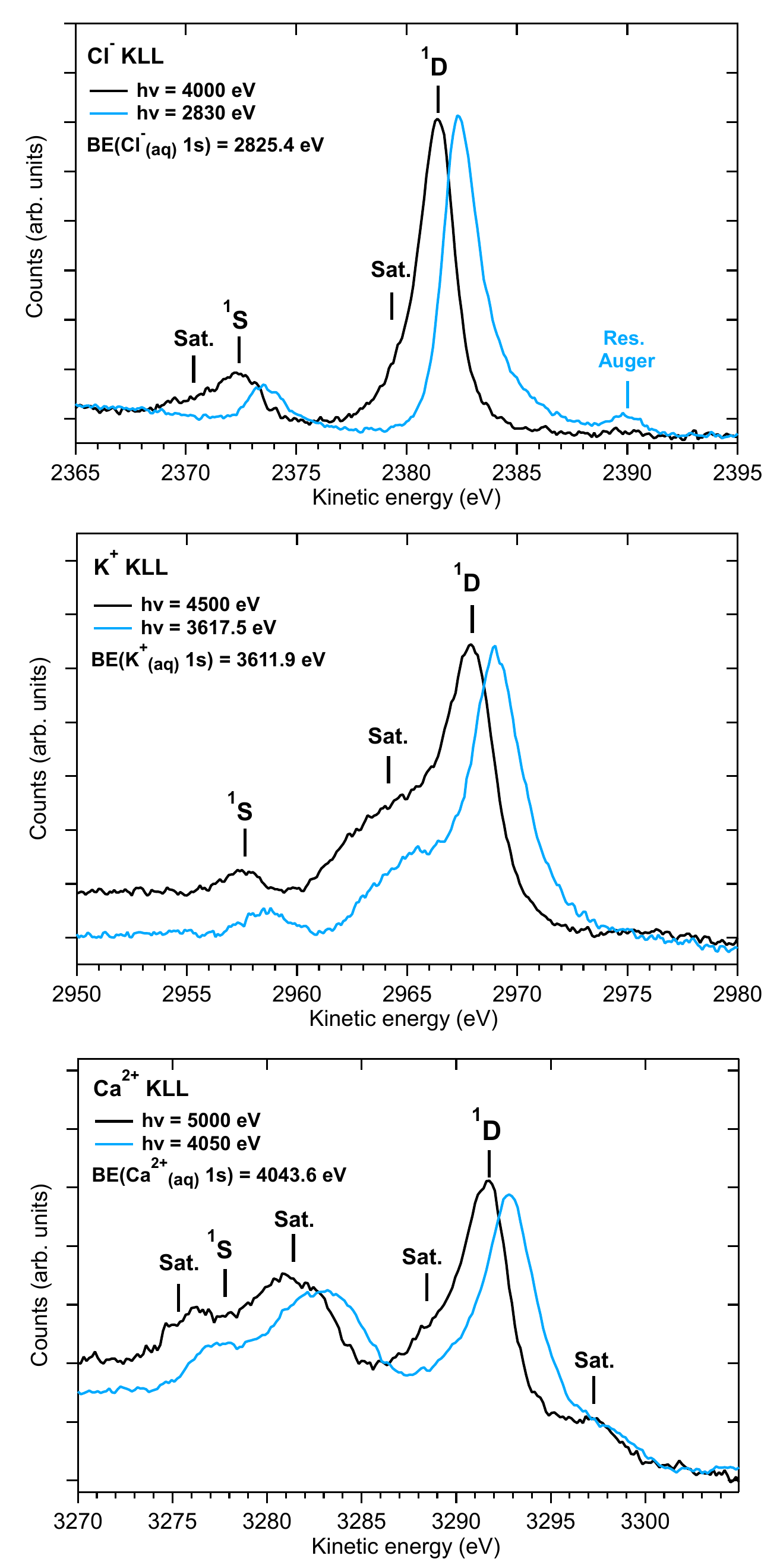}
    \caption{KL$_{2,3}$L$_{2,3}$ Auger spectra of Cl$^-$, K$^+$ and Ca$^{2+}$ in water. The black traces are spectra measured well above threshold ($>$800 eV) in conditions where PCI effects are negligible. The blue traces are spectra measured close to threshold ($\sim$ 6 eV above). The binding energy (BE) of the respective 1s shells is also indicated in the legend.}
    \label{ClKCa_spectra}
\end{figure}

In fig. \ref{ClKCa_spectra}, the KL$_{2,3}$L$_{2,3}$ (hereafter abbreviated KLL) Auger spectra of Cl$^-_{aq}$, K$^+_{aq}$ and Ca$^{2+}_{aq}$ are displayed. For each case the black trace shows the spectrum well above threshold ($>$800 eV), corresponding to conditions where PCI effects are negligible and therefore close to the PCI-free case. Similar spectra were already presented in refs. \cite{ceolin2017,ceolin25}. The spectrum of K$^+$ is characterized by a strong satellite peak on the left side of the main $^1$D peak, which has been attributed to charge transfer states \cite{ceolin2017}. The spectrum of Ca$^{2+}$ contains many strong satellite contributions which are discussed in detail in ref. \cite{ceolin25}. 

The spectrum of Cl$^-$ is the most simple one, although one can see some satellite contributions as well, as both the $^1$D and $^1$S peaks present an asymmetry to the low-KE side. This feature was not previously reported in C\'eolin et al. \cite{ceolin2017}. Its attribution is unclear at the moment, but we assume its behaviour with respect to PCI will be the same as the main peak. It does however complicate the analysis of the shape distortion and maxima shifts induced by PCI, as one needs to account for this initial asymmetric shape. We will nonetheless particularly focus on Cl$^-$ throughout the paper as it exhibits the least complicated shape. 

In fig. \ref{ClKCa_spectra} the blue traces correspond to KLL Auger spectra for the three ions approximately 6 eV above threshold. The binding energy of the respective 1s shells are displayed in the figure. A change of spectral shape is observed, which is most visible in the Cl$^-$ case where the peak is now asymmetric to the high-KE side. The peak maximum is in each case shifted to higher KE by more than 1 eV. Both the shift and distortion are the characteristic hallmarks of PCI. 

\begin{figure}
    \includegraphics[trim={0cm 0cm 0cm 0cm},clip,width=\linewidth]{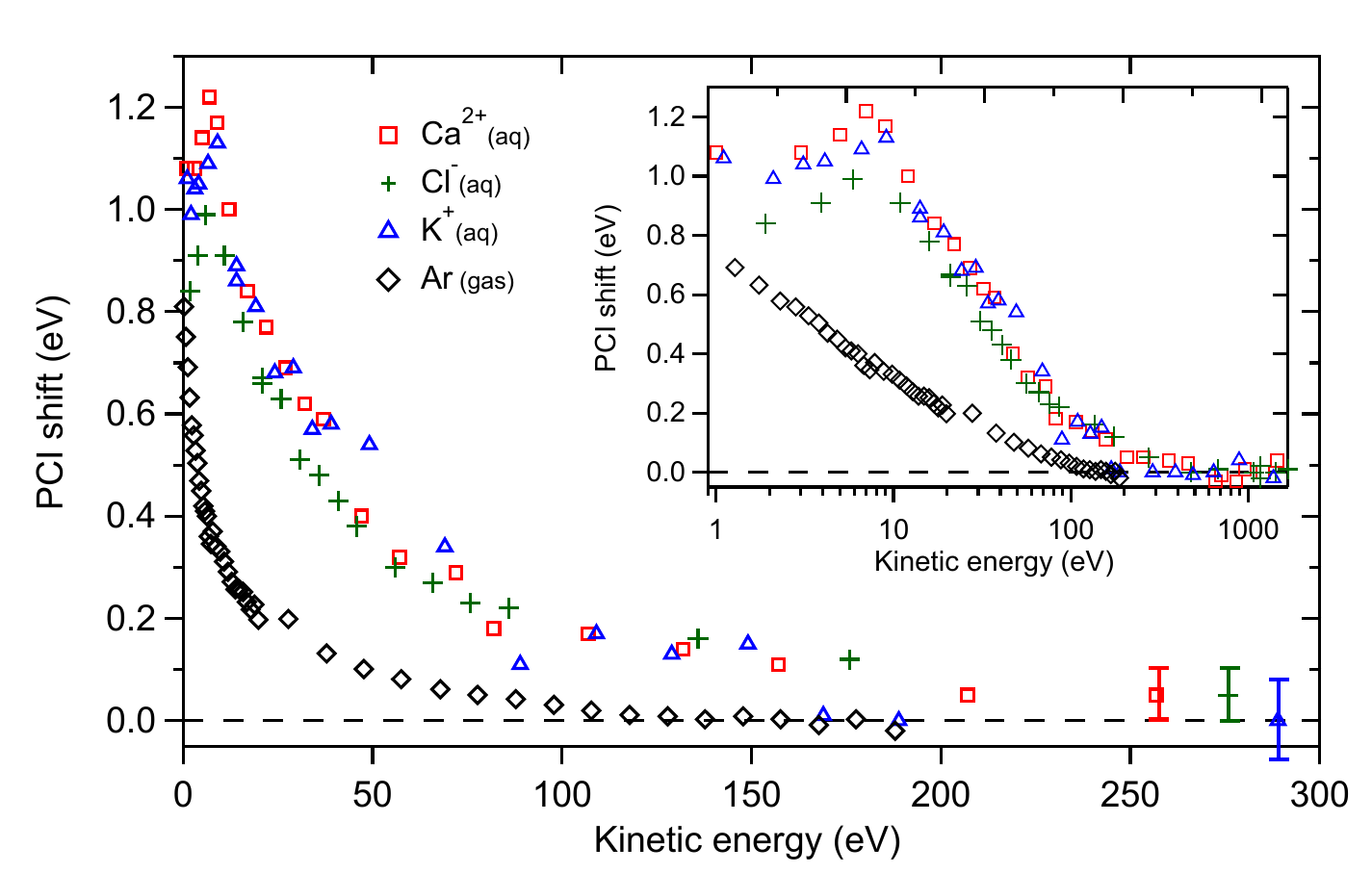}
    \caption{PCI shift (displacement of the peak maximum with respect to its value well-above threshold) for aqueous Cl$^-$, K$^+$ and Ca$^{2+}$. The error bars are only shown on the data points on right-most side of the figure for clarity. The inset is the same plot on a different range, in log scale. The gas phase Argon values from ref. \cite{guillemin2015} are also shown for comparison.}
    \label{PCI_threeions}
\end{figure}

We have investigated how the PCI effect evolves with kinetic energy. In fig. \ref{PCI_threeions} we plot for the three ions the PCI shift, which is the shift of the peak maximum to higher KE compared to its PCI-free position well-above threshold. For comparison we also display the PCI shift measured for the isoelectronic gas-phase Ar by Guillemin et al. \cite{guillemin2015}. While all three ions exhibit roughly similar behaviours, their PCI-shift curves are significantly different from the gas phase one. Similar observations were made by Velasquez et al. \cite{velasquez2023} when comparing gas-phase thiophene and condensed-phase polythiophene. They concluded that the PCI effect is stronger in condensed phase than in gas phase, however comparison of PCI shifts are delicate, since as we mentioned previously it is strongly dependent on the PCI-free peak broadening. In the case of condensed phase spectra, the dominant broadening contribution is due to configurational broadening and peaks are therefore much wider than in the gas phase. In the Appendix, we show that actually in the present case, the PCI shifts are close to what is predicted by the free-atom theory (eq. \ref{straten_spec}) when the PCI-free broadening is taken into account - therefore the intrinsic PCI effect is (fortuitously) roughly the same in the gas and condensed phase. 

One noticeable difference between gas and condensed phase is that in the gas phase, the PCI shift monotonously increases as eKE becomes closer to threshold, while there is a break in trend around 6-10 eV for the solvated ions. We discuss this point later. 

The influence of PCI is marginally stronger for Ca$^{2+}$ than for K$^+$, itself higher than Cl$^-$, which can be attributed to the shorter core-hole lifetime as Z increases, as the difference in kinetic energy of the Auger electrons has very little influence. Note that the K$^+$ dataset exhibits a non-monotonous behavior which we attribute to issues of stability of the kinetic energy scale during the measurements. We observed this behavior on several occasions. The estimated uncertainty on datasets where this occurs is consequently higher. 

\begin{figure}
    \includegraphics[trim={0cm 0cm 0cm 0cm},clip,width=\linewidth]{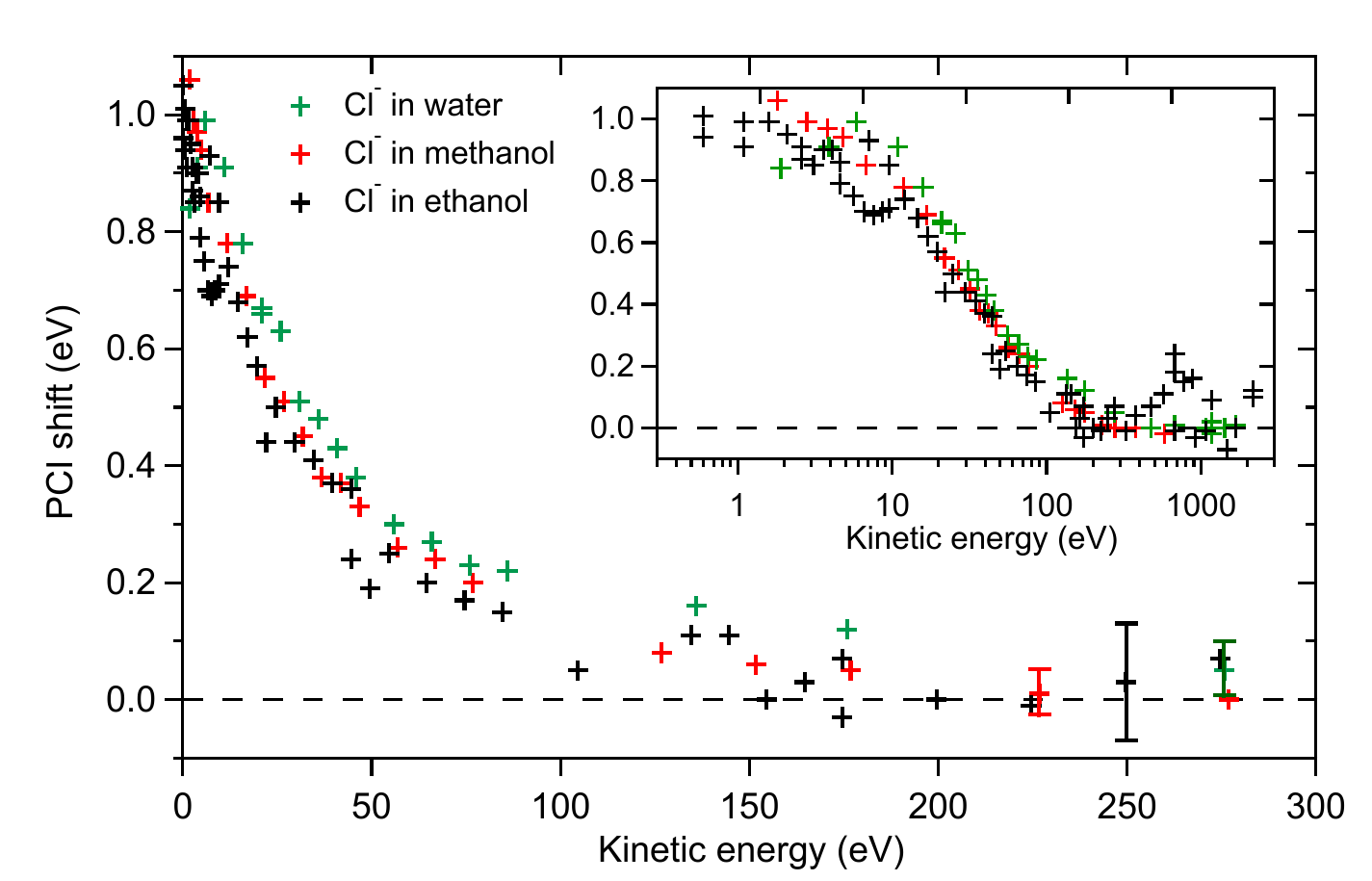}
    \caption{PCI shift (displacement of the peak maximum with respect to its value well-above threshold) for Cl$^-$ solvated in water, methanol and ethanol. The error bars are only shown on the data points on right-most side of the figure for clarity. The inset is the same plot on a different range, in log scale.}
    \label{Cl_solvents}
\end{figure}

We also investigated whether the solvent makes a difference to the PCI effect. In fig. \ref{Cl_solvents} we show the PCI shift as a function of KE for Cl$^-$ in three different solvents: water, methanol and ethanol. The concentration in Cl$^-$ ions is the same each time. The ethanol dataset also exhibits some instabilities of the kinetic energy scale. The PCI shifts seem slightly higher in water than in methanol and ethanol, however, considering the uncertainty on the measurements no strong claim can be made. 

\subsection{Comparison with theory}

In the theoretical framework described in the Methods, PCI in the condensed phase is explained by the free-atom theory modified by the effects of dielectric screening and electron scattering. In this framework, the fact that we obtain similar PCI shifts in gas phase Ar and for the solvated isoelectronic ions (above $\sim$25 eV) once the PCI-free broadening is taken into account (see Appendix) suggests that the effects of scattering, which should increase PCI, and of screening which should reduce it, are of the same order of magnitude and happen to compensate.

Within this framework we can also discuss the fact that no major differences are observed between the three different ions Cl$^-$, K$^+$ and Ca$^{2+}$, or between the three different solvents. The nature of the ion does not influence either screening or scattering, therefore this first result is not surprising. Regarding the solvent, there are large differences in the static dielectric constant of the three solvents but only minor differences between the optical dielectric constant, which as we argued previously is the quantity to consider here. The scattering properties of the three solvents are also certainly different, but not sufficiently to observe a major difference in PCI: the three solvents have similar densities and are composed of light atoms (carbon and oxygen). 

We can also discuss the observed noteworthy feature of the PCI curves mentioned above, which is a flattening of the PCI shift below a threshold around 6-10 eV. This behavior is qualitatively different from what is observed and predicted for free atoms. The 6-10 eV region corresponds to the threshold where electronic scattering channels turn on, and is therefore a point where the scattering behavior changes, which is the most likely explanation for this broken trend. A recent investigation explored this transition regime and its consequences for photoemission spectroscopy close to the threshold in liquids \cite{malerz2021a}. We remain however at a qualitative level concerning this feature, because the simulations we performed for quantitative comparisons cannot handle low-energy behavior properly. 

At higher ($>$ 15 eV) eKE we ran Monte-Carlo simulations and compared the results with the experiment and other theoretical approaches. Comparison of the experimental results with theory requires taking into account the broadening that is not related to the core-hole lifetime, as well as, in the present case, the fact that there are several contributions to the experimental peak. For this reason we only focus on comparison with the Cl$^-$ results, where the main peak shape is the most simple. To account for the peak asymmetry, which is presumably related to satellite features as discussed earlier, this main peak can be deconvolved using two Voigt profiles with a fixed Lorentzian width set to 0.64 eV (corresponding to the lifetime of the Cl 1s core-hole \cite{Fuggle92}). Fitting the spectra well-above threshold we find that the peak can be well-represented using a main component with a Gaussian FWHM of 1.27 eV and a second component, shifted by -0.95 eV, with a Gaussian FWHM of 3.13 eV and an amplitude of 0.58 that of the main component. We convolve all theoretical and simulated spectra with a peak shape composed of two Gaussian profiles with the parameters thus determined. 

\begin{figure}
    \includegraphics[trim={0cm 0cm 0cm 0cm},clip,width=\linewidth]{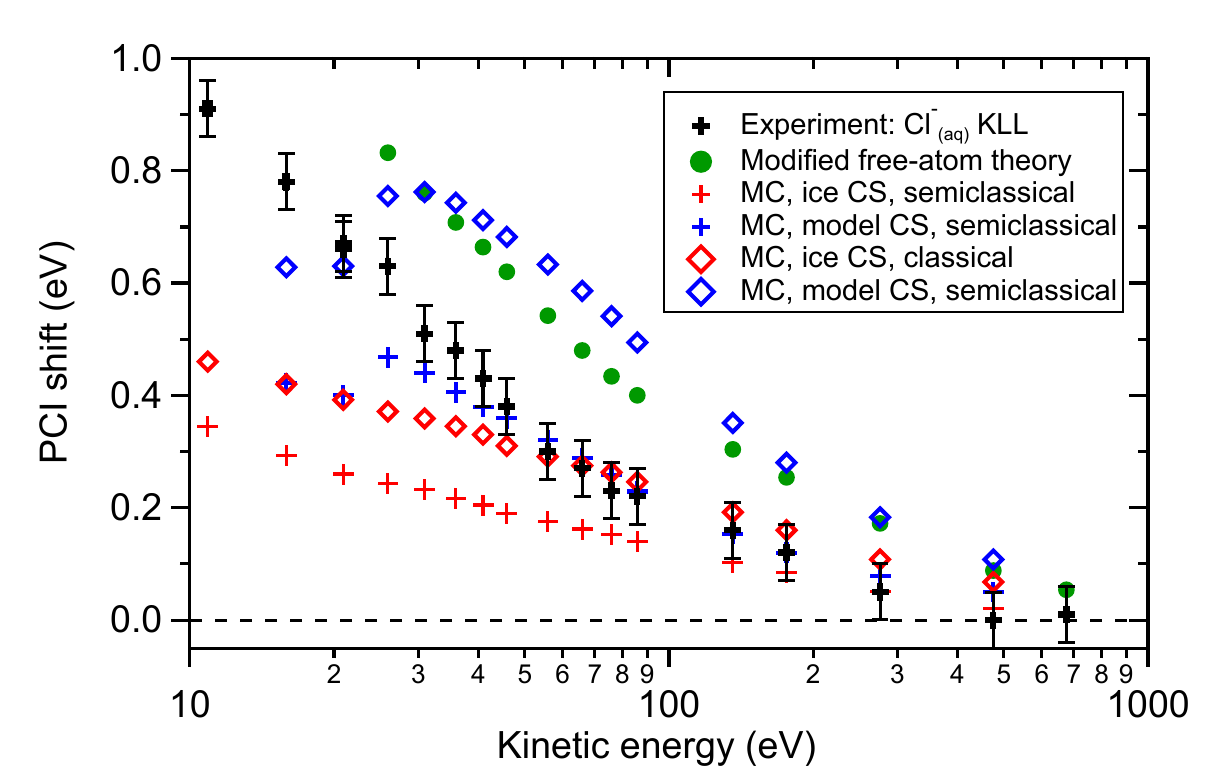}
    \caption{Comparison of the experimental PCI shift for aqueous Cl$^-$ (black crosses) with different theoretical results. The green dots correspond to the modified free-atom theory, eq. \ref{modif_max}. The other points correspond to the results of the Monte-Carlo (MC) simulations. Results obtained with the purely classical approach are shown with crosses and those obtained by mixing the simulation results with the semiclassical free-atom theory integral are shown with diamonds. In red, the results of the simulation using the "ice cross-section (CS)" parameter set and in blue those using the "model cross-sections (CS)" parameter set.}
    \label{Cl_exp_theory}
\end{figure}

In fig. \ref{Cl_exp_theory} we display a comparison of the experimental PCI shift for Cl$^-$ in water with several theoretical results. We display results of the modified free-atom theory proposed in Velasquez et al. \cite{velasquez2023}, i.e. equation \ref{modif_max}, using EAL values for water from ref. \cite{thurmer2013a}. We also show the values computed from the results of our Monte-Carlo simulations. Simulations were performed using the two different "ice cross-sections" parameter set and "model cross-sections" parameter set (see Methods and Appendix). PCI shifts and curves were computed from the results of the electron transport simulations with two different approaches described in the Methods: one purely classical, and one using the simulation output to compute the integral of the semiclassical free-atom PCI theory. The results of the four different combinations are displayed in the figure. 

We can observe that the modified free-atom theory and the "model CS" simulation with classical computation of the PCI shift both overestimate the PCI shift on the whole range, while the "ice CS" simulation with semiclassical computation underestimates it. The "model CS" simulation with semiclassical computation, and the "ice CS" with classical computation, provide a fair agreement, although not on the full kinetic energy range. At low kinetic energy (below $\sim$ 25-30 eV), such a disagreement is not surprising as the cross-sections are not well constrained. This is especially the case for the "model CS" parameter set, for which the analytical extrapolations of the parameters are invalid at too low energies. We can nonetheless observe that there is also a poorer agreement at high eKE, as experiment shows an already vanishing PCI shift at $\sim$ 475 eV above threshold (within the error bars), while the different theoretical results predict non-negligible shifts of the order of 0.1 eV.  

\begin{figure}
    \includegraphics[trim={0cm 0cm 0cm 0cm},clip,width=\linewidth]{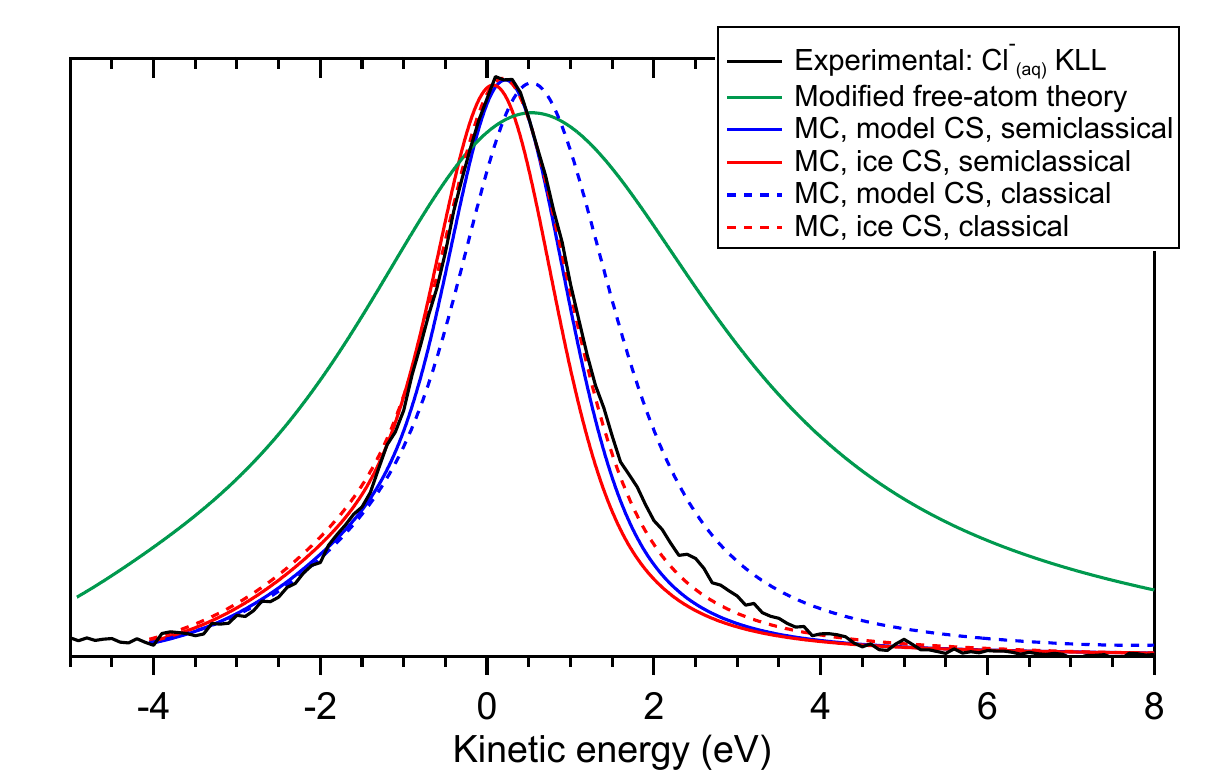}
    \caption{Comparison of the experimental PCI lineshape at eKE = 55.9 eV for aqueous Cl$^-$ (black line) with different theoretical results. The green line corresponds to the modified free-atom theory, eq. \ref{modif_spec}. The other lines correspond to the results of the Monte-Carlo (MC) simulations. Results obtained with the purely classical approach are shown in dashed lines and those obtained by mixing the simulation results with the semiclassical free-atom theory integral are shown with full lines. In red, the results of the simulation using the "ice cross-section (CS)" parameter set and in blue those using the "model cross-sections (CS)" parameter set.}
    \label{Cl_exptheory_shape}
\end{figure}

Before discussing further the merits of the different theoretical approaches, it is also instructive to look at the predicted lineshape instead of only the shifts. In fig. \ref{Cl_exptheory_shape}, we selected the experimental lineshape at eKE = 55.9 eV, an energy high enough that the simulations use correct parameters, but low enough to observe some distortions relative to the PCI-free lineshape. Here it is clear that the green line, corresponding to the modified free-atom theory (eq. \ref{modif_spec}), is much too broad and does not resemble the experiment. The other lineshapes agree better, and the ice CS / classical and model CS / semiclassical combinations both predict closely fitting lineshapes. 

\section{Discussion}

PCI is an object of fundamental atomic physics and has shown very few more practical applications, although it has been used before as a "clocking" tool to investigate the time dependence of other fundamental phenomena \cite{trinter2013a}. Advances in the experimental and theoretical knowledge of PCI in free atoms and molecules have been mainly driven by progresses in experimental techniques allowing to investigate the finer details of the process and a need for theory to explain these finer details. In the continuity of this, liquid-phase electron spectroscopy in the tender X-ray range has allowed here to demonstrate PCI in liquids. We attempt to further contribute to the theoretical description of this phenomenon in condensed phase, and we will therefore now discuss the merits and limits of this description which definitely calls for further work before reaching a satisfying conclusion. We will then also discuss connections with other scientific questions which are slightly more practical, although they remain fundamental. 

\subsection{Merits and limits of the theoretical description}

While a clearly better agreement with experiment can be obtained with the Monte-Carlo simulation approach to scattering, compared with the free-atom theory modified by simple ad-hoc parameters, it is necessary to discuss the validity of the methods employed. In particular, both the classical and "semiclassical" approaches used here have drawbacks, and the agreement with experiment does not necessarily validate either. 

First, we should discuss the meaning of the discrepancies observed between the two different cross-section parameter sets that we have used. The ice CS and model CS sets can be considered respectively as the lower-end and higher-end of the possible values for the scattering cross-sections. This topic is discussed in more details in the Appendix. The large difference observed in the PCI shift curve between the two parameter sets therefore illustrates the high sensitivity of the results to the choice of scattering parameters. Since there is no definitive consensus on which of these parameter sets is more appropriate, it is difficult to make a more firm conclusion on the agreement between these simulation results and the experiment. As we argue later in the article, the PCI experimental results could be used as an argument in favor of the validity of one or the other of these parameter sets, but this is made difficult by the uncertainty remaining on the choice of theoretical approach. 

Indeed, we have used two different approaches to compute the PCI shifts and lineshapes from the MC simulation results, and one yields a good agreement using the ice CS parameter set while the other does so with the model CS parameter set. The classical approach is the most intuitive one. Its main strong point is consistency with the choice of handling electron scattering using electron transport MC simulations, which by design treats the problem in a classical way, with classical trajectories for the electrons. Its drawback is that PCI is initially a fundamentally quantum problem: a classical PCI theory would not correctly predict the wealth of gas-phase PCI data, as the interference terms that contribute in the calculation of wave function overlaps do not exist classically. Nonetheless, one argument in favor of the classical treatment despite this shortcoming is that in condensed phase, scattering will induce a decoherence of the outgoing electron wavefunctions, which can justify a classical treatment. It is however not clear whether we can assume a fully decoherent process. 

This is the reason why we have considered the second approach: starting from integral \ref{int_PCI}, which is a result of the quantum treatment of PCI, and inserting the part that cannot be easily treated at the quantum level, i.e. the electron scattering, from the MC simulations. The merits and limits of this approach are opposite to the classical one: we keep the initially quantum treatment of the free-atom PCI theory, but doing so we introduce the results of a purely classical approach of scattering, and justifying in a rigorous way the mixing of the two is difficult. One argument in favor of doing so is that, although the free-atom PCI theory can be treated in a fully quantum way, it is possible to derive the main results of the eikonal, semiclassical approximation of the theory by introducing classical heuristics into the quantum integral of equation \ref{int_PCI}, in a way similar to what we do here. In conclusion, we do not believe that there is a definitive argument in favor of one approach over the other, which is why we choose to present both, and our experimental results alone cannot decide between them because of the large range of possible scattering parameters. Further theoretical investigations and/or experiments will be necessary to clarify these points. 

Aside from this fundamental question on whether condensed-phase PCI should be treated in a classical or semiclassical way (and how), there are also other possible shortcomings of the current work. Many refinings of the MC simulation code used here would be possible, as we employed a rather simple code far from the sophistication reached by other models in the literature. Some questions also remain regarding screening, in particular which exact value of the optical dielectric permittivity should be considered, and whether this simple approach to screening is accurate. It is known that dielectric screening can be vastly different on a microscopic scale where the structure of the solvent plays a role \cite{vatin2021}, although the length scale where the usual dielectric constant cannot be used has been suggested to be of the order of 1.5-2 nm, while the PCI interaction typically occurs at larger distances except for the slowest electrons. Whether the core-hole lifetime can truly be approximated by the atomic value for solvated ions is another outstanding question. 

\subsection{Consequences of PCI in condensed-phase electron spectroscopy} 

One possible consequence of PCI in condensed phase that should be discussed bears on the precision of measured kinetic (and thus binding) energies in electron spectroscopy. XPS on liquids is commonly conducted at photoelectron kinetic energies around 100 eV, where the cross-section is high and the surface sensitivity at a maximum. This is less true of XPS studies of solids and surfaces, but remains potentially relevant. PCI shifts could potentially introduce errors in the determination of kinetic energies in such measurements. Here we measure shifts of the order of 200 meV on the Auger peak around 100 eV above threshold, which is definitely significant. 

However several factors will mitigate this. First, the overwhelming majority of liquid-phase XPS measurements are conducted on shallow core-levels ($<$ 1000 eV binding energy) for which the Auger lifetime is much longer (e.g. 4 fs for O 1s) and the Auger electron slower, reducing the magnitude of PCI. Second, the PCI shift is observed here on the Auger peak, but there is no symmetry with the effect that would be observed on the photoelectron, for several reasons. If one uses the photoelectron as an observable, then by definition only the photoelectrons that have not been inelastically scattered are probed, which reduces the magnitude of PCI. Additionally, the probing depth is also different and the much higher surface sensitivity of the photoelectron measurement will further reduce the effects of scattering. 

If simulations are ran using parameters more typical of usual liquid-phase XPS measurements, for instance those for O 1s photoionization (0.17 eV core-hole lifetime \cite{nicolas2012}, 500 eV Auger kinetic energy \cite{thurmer2013}, a corresponding EAL of 30 \AA~for the Auger electron or 10 \AA~for a 100 eV electron \cite{thurmer2013a}), the PCI shifts for eKE = 100 eV are of the order of 10-20 meV. This is a rather minute shift to discern on broad features (the O 1s photoelectron peak is 1.5 eV wide, which is not uncommon for liquid-phase peaks). In their recent investigation of the exact first vertical ionization energy (VIE) of liquid water, Thürmer et al.\cite{thurmer2021} reach an overall precision of $\pm$30 meV on the water 1b$_1$ peak (which is unaffected by PCI effects, being a valence level). Such a small shift could therefore in principle be resolved by very careful experiments, but it would have no consequences in most practical uses of liquid-phase XPS.  

\subsection{PCI as a test of scattering cross-sections} 

Another interesting axis of investigation concerns the scattering cross-sections of aqueous solutions. Turning the problem around, if we assume to have reached a sufficiently satisfying theoretical model of PCI in the condensed phase (which may not be the case currently), it would be possible to use experimental PCI measurements as a test of the scattering cross-section parameter sets used as input of the simulations. Currently, there is a large uncertainty on the inelastic and elastic scattering cross-sections of electrons in liquid water - both integral and differential, although the former has been more investigated. Direct, precise experimental measurements of these quantities seem elusive. Experiments on liquid water have mainly attempted to indirectly derive one specific quantity, the effective attenuation length (EAL) \cite{ottosson2010,thurmer2013,suzuki2014,schild2020}. 

The two other sources of experimental data which can and are used in electron transport and radiolysis simulations in water are gas-phase data, for which there are several published works, and ice data, which is mainly the work of Michaud and Sanche on amorphous ice films \cite{michaud2003}. Gas-phase data has been largely used to build semi-empirical models or validate theoretical models of scattering, for lack of liquid-phase data. As discussed in the Appendix, there is however a large discrepancy between the gas and ice cross-sections, and it is questionable whether gas phase data can truly be used for liquid water modeling. 

One possibly fruitful approach to determine appropriate scattering parameters for liquid water has been opened by Signorell \cite{signorell2020}, who used the ice scattering CS of Michaud and Sanche to model different photoemission experiments with success, within the (nonetheless large) uncertainties. Since direct measurements on liquids using a single method seems elusive, using instead an array of experiments as large as possible as a benchmark for models could provide a mean to refine and validate cross-section parameter sets. PCI data is a useful addition to such an array of experiments, especially since it is also sensitive to the differential cross-sections, whereas other experiments are mostly sensitive to the integral cross-sections only. The recent use of high-harmonic generation (HHG) in liquid water data to obtain information on electron scattering \cite{mondal2023} is also a step towards this diversification of experiments that could be used as benchmarks.
 
\section{Conclusion}

We have observed the PCI effect on the KLL Auger decay of three isoelectronic ions, Cl$^-$, K$^+$ and Ca$^{2+}$, solvated in water, as well as for Cl$^-$ solvated in methanol and ethanol. The magnitude of the effect is similar in all cases, with differences that barely exceed the uncertainties. These small differences can be attributed to the slightly different core-hole lifetimes of the ions and slightly different scattering and dielectric properties of the solvent. While the PCI effect appears to be stronger than what has been measured for the isoelectronic argon atom in the gas phase, this turns out to be mostly an effect of the much larger PCI-free width of the Auger peak. PCI in the condensed phase differs from the gas-phase by two counter-acting effects: screening of the coulombic interactions, which reduces the magnitude of PCI, and electron scattering, which increases it. While in small free clusters and metals the former effect dominates and PCI is quenched, in bulk dielectric media scattering plays an important role, and here fortuitously compensates roughly the effect of screening. 

To model theoretically our experimental results, we first considered the model proposed in Velasquez et al., a modification of the free-atom theory with additional parameters to account for screening and scattering. Screening in a dielectric medium is relatively easy to take into account, by dividing the interaction energy between the electrons by the dielectric permittivity, although which dielectric permittivity should be considered is not completely straightforward. We discussed that the optical permittivity is a reasonable choice, considering the timescale of PCI. Scattering, on the other hand, is taken into account only in a qualitative way in this theory. Consequently the agreement with experiment is not excellent for the PCI shift, and clearly off for the PCI lineshape. 

We proposed an alternative approach to tackle electron scattering, which consists in performing electron transport Monte-Carlo simulations. This approach yields a better agreement, especially on the lineshape, although the lack of strong constraints on the scattering parameters governing the results (the integral and differential cross-sections for elastic and inelastic scattering) leads to rather large uncertainties. Furthermore, we have explored two different approaches to compute the PCI shifts and lineshapes from the results of the electron transport simulations. One is purely classical and the other mixes the simulation results with the quantum integral at the basis of the semiclassical free-atom PCI theory. Both have merits and drawbacks and experiment alone cannot decide on which is better. 

We suggest that constraining the electron scattering parameters of liquid water, a long sought-after goal, will require considering a diverse set of experiments (photoemission, photoelectron angular distributions, high-harmonic generation...), of which PCI data can be part of, if the correct theoretical approach can be settled on. 

\begin{acknowledgments}

The authors would like to thank L. Gerchikov, S. Sheinerman and P. Lablanquie for fruitful discussions. The authors thank the SOLEIL synchrotron facility for provision of beamtime under projects 20181191, 20191240 and 99190147, and the GALAXIES beamline team for support. R.D. thanks R. Signorell for a discussion on electron transport models, and S. Thürmer for providing and constantly improving his Igor Pro tools. 

\end{acknowledgments} 

\section*{Data Availability}

The raw data that support the findings of this article are openly available \footnote{Data repository available at doi:10.5281/zenodo.15973036}

%


\appendix

\section{Details on the Monte-Carlo Simulations}

We will describe here the principle of the MC simulations that were performed, as only a brief description was given in the main text. 

As described in the main text, the PCI lineshape was computed using two different approaches, one classical and one "semi-classical". In both cases, the input from the MC simulation that serves to take into account the effects of scattering is the quantity S(t), which is the energy exchanged between the two electrons. We further assume that S(t) = 1 / $\epsilon$r(t) where r(t) is the crossing radius where the Auger electron overtakes the photoelectron. In the free atom theory under the eikonal approximation, r(t) could be determined analytically as the electrons have rectilinear uniform motions. If we are to take into account scattering, the electrons will be deviated from the rectilinear trajectory by elastic scattering and decelerated by energy losses from inelastic scattering. The difficulty of providing an analytical treatment of these phenomena, even in an average way, is what leads to the use of MC simulations. What is extracted from the simulation is an (average) value of r for a given t (time at which the Auger electron is emitted) and for given scattering conditions.

\subsection{Electron transport}

As scattering is a stochastic process, r(t) will vary from event to event and we will consider a statistical average of this quantity. Scattering is governed by several parameters which are complex to model analytically, therefore Monte-Carlo simulations are an appropriate way to determine this average. In these simulations, we only simulate scattering of the photoelectron. The Auger electron energy is the experimental observable: it means that we observe only Auger electrons which have not lost any kinetic energy to inelastic scattering. We also neglect elastic scattering as Auger electrons are assumed to have an initially isotropic angular distribution. 

The simulation follows the spatial evolution of the photoelectron with time (t = 0 being the photoelectron emission time). To determine when a scattering event happens, a scattering mean free time is calculated:

\begin{equation}
    \mathrm{t_{MFP}} = \frac{1}{\rho (\sigma_{\mathrm{E}} + \sigma_{\mathrm{I}}) \mathrm{v_{ph}}}
\end{equation}

Where $\sigma_{\mathrm{E}}$ and $\sigma_{\mathrm{I}}$ are respectively the elastic and inelastic integral scattering cross-sections, $\rho$ the molecular density of water (taken as 3.343 $\times$ 10$^{22}$ cm$^{-3}$ \cite{signorell2020}) and v$_{\mathrm{ph}}$ the speed of the photoelectron. In principle v$_{\mathrm{ph}}$ is not constant between two scattering events as the electron decelerates in the Coulomb field of the ion, but we neglect the deceleration and take it as constant. The time between two scattering events is drawn randomly according to $\mathrm{t_{step}}$ = - $\mathrm{t_{MFP}} \times$ln(rand()) where rand() is a random number in ]0,1]. Between scattering events, the trajectory of the electrons are calculated by solving the equations of movement with two- or three-body Coulomb interactions depending on whether the Auger electron has been emitted or not. The main purpose is to take into account deceleration of the electrons in the Coulomb field of the ion; the electron-electron interaction has very little effect in the vast majority of trajectories. To save computation time, these interactions are turned off once the electrons are sufficiently far away ($>$ 40 a.u.) from the ion.

When a scattering event occurs, first it is determined randomly based on the ratio of the cross-sections whether it is elastic or inelastic. If the event is elastic, the trajectory of the electron is deviated by an angle $\theta$ drawn according to a probability distribution, the differential  cross-section (DCS) for elastic scattering with respect to angle. The elastic DCS d$\sigma_\mathrm{E}$/d$\theta$ is therefore an additional parameter of the simulation. If the event is inelastic, the electron is not deviated but loses energy. This energy loss is again drawn according to a probability distribution, the differential cross-section for inelastic scattering with respect to energy loss, d$\sigma_\mathrm{I}$/dE. 

Whenever the photoelectron energy changes (due to inelastic scattering or deceleration in the ion field) the relevant energy-dependent parameters of the simulation ($\sigma_\mathrm{I}$, $\sigma_\mathrm{E}$, d$\sigma_\mathrm{I}$/dE,  d$\sigma_\mathrm{E}$/d$\theta$ and the related quantities such as $ \mathrm{t_{MFP}}$) are updated. 

\subsection{Ice cross-section parameter set}

The ice cross-section parameter set is called thus because it is based on the scattering cross-sections determined by Michaud and Sanche \cite{michaud2003} using high-resolution electron energy loss spectroscopy (HREELS) on amorphous ice films grown on a substrate under vacuum and a detailed analytical modelling of the energy loss spectra they obtained. Signorell \cite{signorell2020} recently showed that these cross-sections could be used without further modifications in an electron transport model to reproduce within the (nonetheless large) uncertainties three separate liquid photoemission experiments, validating their usefulness for the description of electron transport in water. 

Michaud and Sanche give detailed cross-sections for various scattering channels, including true elastic scattering, individual vibrational and librational scattering channels, and an "other" inelastic scattering channel which subsumes electronic inelastic scattering (dominant a high kinetic energies) and a few other processes. It is the latter that we use as our integral inelastic scattering cross-section. For energies above 100 eV, we scale the values of Shinotsuka et al. \cite{shinotsuka2017} to those of Michaud and Sanche, similarly to Signorell \cite{signorell2020}.

For elastic scattering, we take the transport cross-section values recommended by Signorell \cite{signorell2020}. The transport cross-section is an effective isotropic scattering cross-section: the corresponding transport mean free path is an estimate of the distance after which an electron has lost memory of its initial direction. Consequently, we use with this transport cross-section an isotropic differential scattering cross-section, i.e. when an elastic event occurs the deviation angle is randomly picked over $\pi$ with equiprobability. This is clearly a shortcoming of this model, as in reality the deviation of the photoelectron would occur by a series of elastic scattering events for which it is more probable to have forward scattering. The spatial trajectories thus generated will not be so accurate. The reason for this choice is that to begin with the analysis of Michaud and Sanche is based on a "two-stream" approximation of scattering using a single parameter describing the forward-backward scattering asymmetry, with a different parameter for each scattering channel. Since we only use a two-channel model here, we avoid complications by opting for a transport integral cross-section and an isotropic DCS.  

For the differential inelastic scattering with respect to energy loss nothing is provided by Michaud and Sanche. We opt instead for an ad hoc modification of the analytical form for energy loss given by Tougaard \cite{tougaard1997}. This analytical form was originally created to provide a convenient way to analyze backgrounds in electron spectroscopy of solids, which are largely due to inelastically scattered electrons. For an insulator, the probability P(E) of losing energy E is: 

\begin{equation}
    P(E) = H(E-E_0) \frac{BE}{(C-E^2)^2+DE^2}
\end{equation}

Where B, C and D are empirical parameters, and E$_0$ is the bandgap energy of the material, thus H(E-E$_0$) with H the Heaviside function characterizes the fact that for insulators there is a minimum electronic energy loss, which in molecular physics terms corresponds to the first excited electronic state. For water, E$_0$ = 7.5 eV. Parameters C and D are set respectively to C = 500 and D = 650 according to a fit of the differential cross-section of ref. \cite{emfietzoglou2003}. Parameter B is unimportant as it is only a scaling parameter, and only the shape of the cross-section is used in the code. In practice, we use the following function:

\begin{equation}
    P(E) = \frac{BE}{(C-E^2)^2+DE^2} \Theta(E-E_0) (1 - \Theta(E - E_{max}) )
\end{equation}

Where $\Theta$ is the logistic sigmoid function, used as a smoothed out version of the Heaviside function. The second $\Theta$ function is a cut-off at high energy, to ensure the energy loss is below the kinetic energy of the photoelectron. This energy loss cross-section is not necessarily realistic, especially at low kinetic energies, as it was mainly parameterized for higher kinetic energies, but it has a convenient analytical expression and a reasonable shape. 

\subsection{Model cross-section parameter set}

The second parameter set we built used instead cross-sections derived from commonly used semi-empirical models of scattering \cite{Plante11}, which are mainly based on measured gas-phase water cross-sections. For the inelastic scattering cross-section, we used the models for water ionization and excitation from respectively Rudd \cite{rudd1990} and Kutcher and Green \cite{kutcher1976}. They give expressions for the differential cross-section for each ionization shell/excited state, which we sum and integrate to obtain the integral cross-section. 

For the integral elastic scattering cross-section, we used the values calculated by Aouchiche et al. \cite{aouchiche2008} for liquid water, while for the differential cross-section we used the parameterization given by Brenner and Zaider \cite{brenner1984}. The treatment of elastic scattering is thus different from the previous parameter set, where we used a transport cross-section with an isotropic differential cross-section. 

\subsection{Differences between the two parameter sets}

A major difference between the two parameter sets used here is that one is based off of the cross-sections determined for ice films, and the other one is mainly based on gas-phase cross-sections. A long discussion is dedicated to the differences between ice and gas cross-sections in Nikjoo et al. \cite{nikjoo2016}, and many insights can also be found in Signorell \cite{signorell2020}. 

\begin{figure}
    \centering
    \includegraphics[trim={0cm 0cm 0cm 0cm},clip,width=\linewidth]{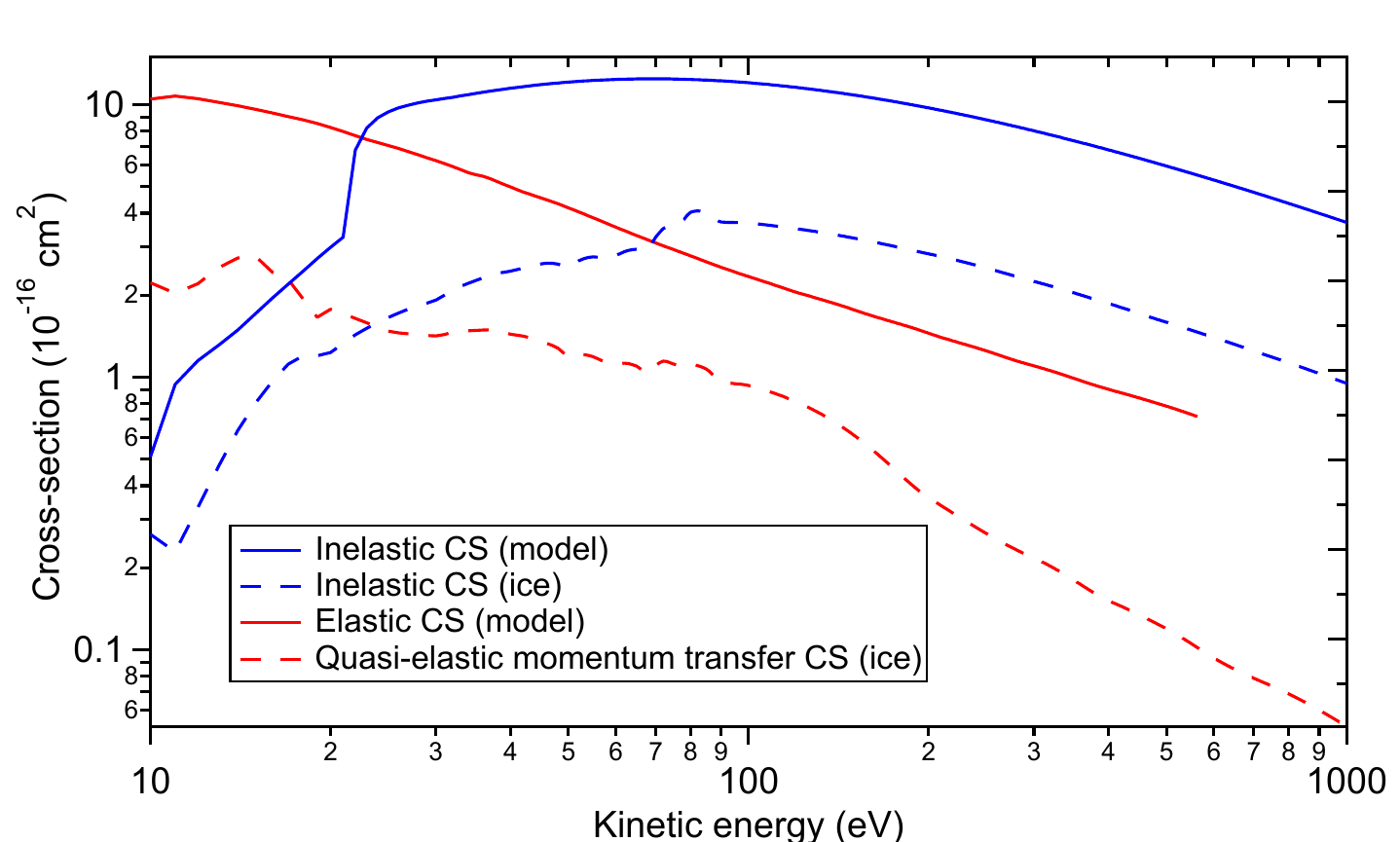}
    \caption{Integral cross-sections for inelastic and elastic scattering used in the two different parameter sets considered in this work ("ice CS" and "model CS"). For the ice CS set, the quasi-elastic momentum transfer CS is considered instead of the elastic CS.}
    \label{SI_CSs}
\end{figure}

In fig. \ref{SI_CSs}, the integral elastic and inelastic cross-sections used in the present work are displayed. One can readily observed that the ice CS and model CS differ by a factor of $\sim$ 5 in the whole kinetic energy range. The difference is actually even more drastic for the elastic CSs, as in the ice case we considered a quasi-elastic momentum transfer CS; the actual elastic CS would be even lower. This large difference is what explains the difference of the results obtained in the MC simulations for the two models.  

\section{Comparison of different theoretical results}

\begin{figure}
    \centering
    \includegraphics[trim={0cm 0cm 0cm 0cm},clip,width=\linewidth]{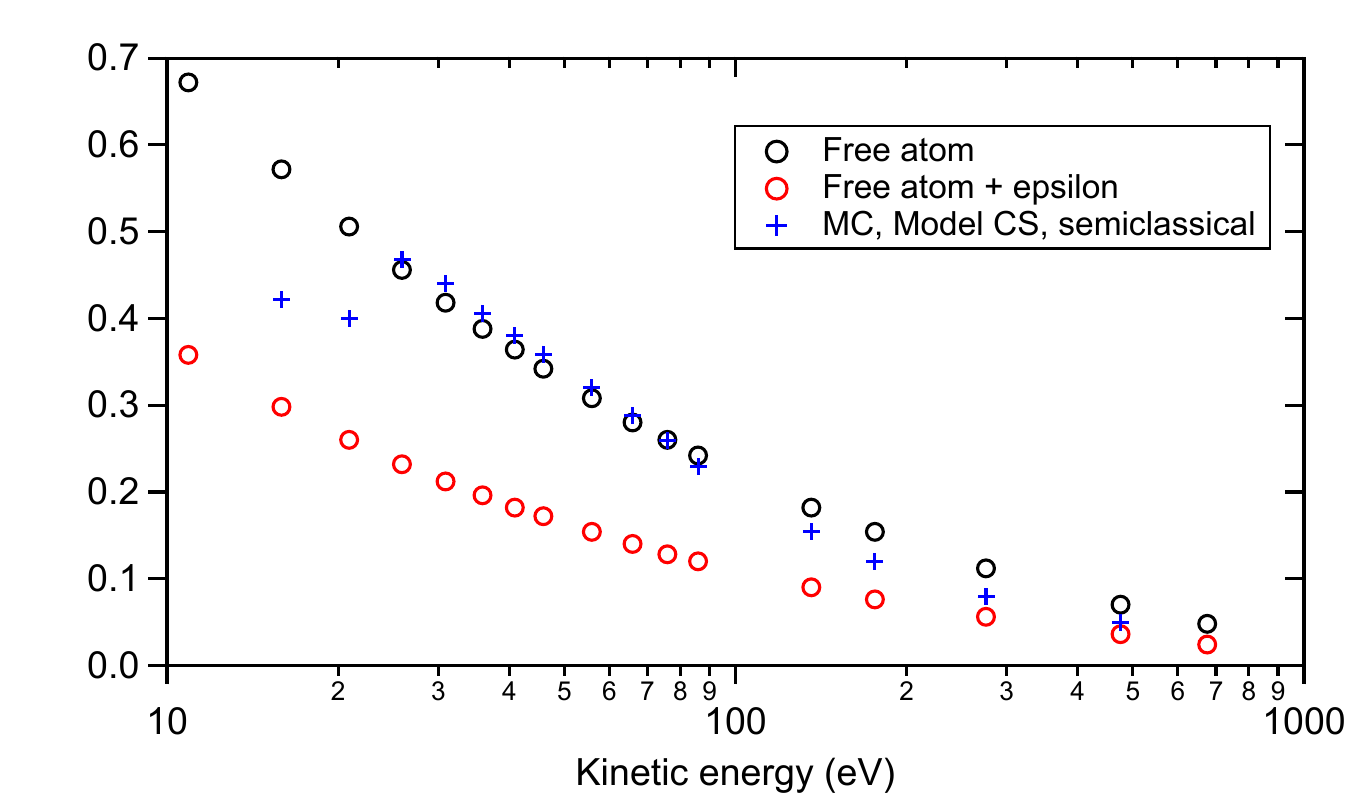}
    \caption{Comparison of PCI shifts for different theoretical models. In all cases the calculated lineshape was convolved with the same convolution shape used to model the experimental results in the main text.}
    \label{SI_theories}
\end{figure}

In fig. \ref{SI_theories} we compare the PCI shift curve obtained from the Monte-Carlo simulation with the "model CS" parameter set and semi-classical computing of the PCI shift with what is predicted by the free atom theory. The model CS / semiclassical combination is one of the two combinations that yielded good agreement with the experimental results, as shown in the main text. We also display the results of the free atom theory modified to account only for dielectric screening - meaning, S(t) in equation 2 of the main text was divided by the optical dielectric constant \textepsilon$_{opt}$ = 1.8. The PCI shifts are obtained by computing the spectral shape in each case, convolving it with the experimental broadening lineshape discussed in the main text, and taking the peak maximum. Adding dielectric screening to the free atom theory reduces the PCI shifts, as expected, while in the full model CS / semiclassical calculation where scattering is also included the PCI shifts become close to the initial free atom results. 

This means that, in the present case, the effects of dielectric screening and scattering compensate roughly. As the model CS / semiclassical computation agrees fairly well with experiment, we must conclude that it is the case also for the experimental situation. This compensation is fortuitous as there is no particular reason for scattering and screening to have opposite effects of similar magnitude on the PCI shifts. Note that in order to reach this conclusion one must be careful to compare only PCI shifts that have been computed after taking into account the different sources of peak broadening. 

\end{document}